# Gravitation and Forces Induced by Zero-Point Phenomena

Charles T. Ridgely
charles@ridgely.ws

A recent proposal asserts that gravitational forces arise due to an interaction between matter and vacuum electromagnetic zero-point radiation. The present analysis demonstrates that forces induced on matter by zero-point radiation arise in addition to gravitational forces. It is argued that zero-point radiation should be red-shifted near large gravitational sources while remaining essentially undetectable within freely falling reference frames. On this basis, an effective weight of an observer stationed near the surface of the Earth is derived for the case when zero-point radiation is present.

## 1. Introduction

In recent times, there have been several attempts to shed greater insight into the origin of inertial and gravitational forces. According to one recent proposal, inertial and gravitational forces alike arise entirely due to an interaction between vacuum electromagnetic zero-point radiation and subatomic particles comprising ordinary matter [1]. For the case of inertia, this proposal suggests that when an object accelerates through the zero-point radiation field (ZPF), pervading all of space, the quarks and electrons comprising the object scatter a portion of the radiation passing through the object. This in turn exerts an electromagnetic drag force on the object, which, according to this ZPF proposal, can be associated with the object's inertia. Additionally, those who support the ZPF proposal also seek to ascribe gravitation entirely to interaction with zero-point radiation [2]. However, such an approach is certainly not without conceptual difficulties.

One persistent source of difficulty surrounds the question of what percentage of the force on an object is actually ZPF induced. This question seems to be hinged on the origin of the rest mass-energy of matter. According to the ZPF proposal, the energy content of ordinary matter is entirely internal kinetic energy due to ZPF-induced jittering motion, or zitterbewegung [1], of quarks and electrons comprising matter. As pointed out in [3], however, ascribing the energy content of subatomic particles entirely to internal kinetic energy effectively neglects the rest mass-energy content of these particles. It is straightforward to see that quarks and electrons each possess charge and spin, and thus are each surrounded by an electromagnetic field. These electromagnetic fields possess energy. Therefore, while ZPF-induced zitterbewegung [1] must certainly give rise to kinetic energy of quarks and electrons, such particles also possess intrinsic quantities of rest mass-energy due at least in part to their electromagnetic fields.

Another seemingly important issue is space-time curvature existing near gravitational sources. The ZPF proposal appears to do away with the notion of space-time curvature, opting in favor of electrodynamic properties of space-time. One example of this can be seen in an attempt to explain the gravitational bending of light near massive bodies by ascribing variable dielectric properties to space [1]. In essence, the ZPF proposal treats space as a polarizable vacuum. As is very well known, however, general relativity predicts that space and time intervals are physically affected by all forms of stress-energy, regardless of origin. With this in mind, it is difficult to imagine how the ZPF proposal, in its present form, can account for the well-documented dilation of time arising near large gravitational sources [4]. In our opinion it is precisely the behavior of space-time predicted by general relativity that gives rise to both inertial and gravitational forces.

In a previous analysis [3], we used special and general relativity to demonstrate that inertia is a purely relativistic phenomenon, arising solely due to space-time anisotropy within accelerating systems. Specifically, for the case of inertia it was shown that an observer who exerts a constant force on an object experiences a reactive force of the form

$$\mathbf{f} = -E \nabla \left( \frac{dt}{d\tau} \right) \qquad (1)$$

wherein $E$ is the total energy content of the object, $d\tau$ is an interval of proper time experienced by the accelerating object, and $dt$ is a corresponding interval of time experienced by the force-producing observer. Using this expression, it was further shown that ZPF-induced forces acting on accelerating matter arise in addition to the intrinsic inertial properties of matter [3]. The present analysis uses the same approach for the case of ZPF-forces induced due to gravitation. Herein it is demonstrated that gravitationally induced ZPF-forces acting on an observer residing near a gravitational source arise in addition to the purely gravitational force arising due to the intrinsic energy content of the observer and the source.

One noteworthy point to notice is that electromagnetic zero-point radiation possesses energy, and hence ought to be subject to space-time geometry just as are other forms of radiation. Thus, one would expect the ZPF to be red-shifted near large gravitational sources. Such a red-shift ought to arise simply because gravitational space-time curvature gives rise to anisotropy in the electromagnetic mode structure of the ZPF. Additionally, electromagnetic zero-point radiation is Lorentz invariant [1], which suggests that zero-point radiation should be essentially uniformly distributed within local freely falling reference frames. Based on this line of reasoning, we suspect that the ZPF should exert some level of force on an observer stationed near a gravitational source, while remaining essentially undetectable to freely falling observers. This observation is used herein to demonstrate that ZPF-induced forces act in addition to gravitational forces.

## 2. Weight due to Gravitational Anisotropy of Zero-Point Radiation

Let an observer of proper mass *m* be stationed a distance *r* from the center of a weak gravitational source such as the Earth. Within the stationary reference frame of the observer, the ZPF is observably red-shifted due to space-time curvature near the Earth. Such a red-shift of the ZPF gives rise to a body force on the observer, which acts in addition to the gravitational force on the observer. According to [3], the total force acting on the observer may be expressed in the form

$$\mathbf{f} = -E'\nabla\left(\frac{dt}{d\tau}\right) \qquad (2)$$

where $E'$ is the total energy content of the observer and $dt/d\tau$ is a scalar function that characterizes the gravitational anisotropy of space-time near the Earth. One important point to notice is that $E'$ represents all forms of energy possessed by the observer, and is not limited to merely the ZPF-induced energy. Thus, the total energy content of the observer must be expressed as $E' = E + E_{ZPF} + U$, where *E* is the intrinsic mass-energy of the observer, $E_{ZPF}$ is the internal ZPF-induced zitterbewegung [1] of the subatomic particles comprising the observer, and *U* includes additional forms of energy that may be possessed by the observer, such as due to the strong and weak forces. Using this expression for the energy content of the observer, Eq. (2) can be recast in the form

$$\mathbf{f} = -(E + E_{ZPF} + U)\nabla\left(\frac{dt}{d\tau}\right) \qquad (3)$$

The sum enclosed within parentheses makes it clear that the interaction between zero-point radiation and the subatomic particles comprising the observer contributes positively to the passive gravitational mass-energy of the observer.

In order to evaluate Eq. (3), an expression for the scalar function $dt/d\tau$ must first be derived. As is very well known, the geometry of space-time exterior to a large spherical body is described by the Schwarzschild coordinate system, having a space-time interval of the form

$$ds^2 = \left(1 - \frac{2GM}{c^2 r}\right)c^2 dt^2 - \frac{dr^2}{\left(1 - \frac{2GM}{c^2 r}\right)} - r^2 d\theta^2 - r^2 \sin^2\theta\, d\phi^2 \qquad (4)$$

in which *G* is the gravitational constant, *M* is the total mass of the body, and $\{r, \theta, \phi\}$ are spherical coordinates exterior to the source. Using Eq. (4) for the case of a stationary observer, the scalar function is easily found to be

$$\frac{dt}{d\tau} = \frac{1}{\sqrt{1 - \frac{2GM}{c^2 r}}} \qquad (5)$$

wherein $d\tau$ is an interval of proper time experienced by the stationary observer and *dt* is an interval of time experienced by a free fall observer whose coordinate origin is momentarily coincident with that of the stationary observer at a time $t = 0$. For the case of small, weakly gravitating sources, Eq. (5) simplifies to an approximate form given by

$$\frac{dt}{d\tau} \approx 1 + \frac{GM}{c^2 r} \qquad (6)$$

This expression holds for the case of small, weakly gravitating sources such as the Earth.

Upon using Eq. (6) and expressing the gradient operator in terms of the Schwarzschild coordinate system, Eq. (3) simplifies to

$$\mathbf{f} = -(E + E_{ZPF} + U)\hat{r}\frac{\partial}{\partial r}\left(1 + \frac{GM}{c^2 r}\right) \qquad (7)$$

where $\hat{r}$ is a unit vector in the radial direction relative to the Earth, and only first order terms have been retained. Carrying out the partial differentiation, and simplifying somewhat, Eq. (7) reduces to

$$\mathbf{f} = -\frac{GMm}{r^2}\hat{r} - \frac{GM}{c^2 r^2}E_{ZPF}\hat{r} - \frac{GM}{c^2 r^2}U\hat{r} \qquad (8)$$

wherein $E = mc^2$ has been used to simplify the first term. Equation (8) can be simplified further upon noticing that according to the ZPF proposal, the energy contributed to the observer due to interaction with zero-point radiation is expressed as [1]

$$E_{ZPF} = V\int \eta(\omega)\frac{\hbar\omega^3}{2\pi^2 c^3}d\omega \qquad (9)$$

In this expression, *V* is the proper volume of the observer and $\eta(\omega)$ is a spectral function that governs the extent to which zero-point radiation actually interacts with the observer. Using Eq. (9) in Eq. (8), and rearranging, leads to

$$\mathbf{f} = -\frac{GMm}{r^2}\hat{r} - \frac{GMV}{c^2 r^2}\hat{r}\int\eta(\omega)\frac{\hbar\omega^3}{2\pi^2 c^3}d\omega - \frac{GM}{c^2 r^2}U\hat{r} \qquad (10)$$

This is the effective weight of an observer stationed at a radial distance *r* from the center of the Earth in the presence of zero-point radiation. The first term is easily identified as the usual Newtonian expression for the gravitational force between two massive bodies. The second term is an additional force on the observer arising due to gravitationally induced scattering of zero-point radiation. The third term arises due to additional forms of energy with which the observer may be endowed, such as due to the strong and weak forces. The force due to zero-point radiation is clearly an additional downward-acting body force that contributes to the measurable weight of the stationary observer.

## 3. Discussion

The present analysis argued that zero-point radiation is *not* the source of gravitation, but rather that ZPF-induced forces comprise additional gravitationally induced forces on matter.

To demonstrate this, an expression for the effective weight of an observer stationed near the Earth in the presence of zero-point radiation was derived. The derivation was performed on the basis that zero-point radiation is red-shifted near gravitational sources while remaining substantially undetectable within the local reference frames of free fall observers. The resulting expression for the effective weight suggests that gravitationally induced ZPF-forces act in addition to gravitational forces, thereby increasing the weights of stationary observers.

It seems that the next natural question to ask is can the ZPF be manipulated in such a manner that ZPF-induced forces act in opposition to gravitational forces? According to [5], both gravitational and inertial forces alike may be affected by altering the electromagnetic mode structure of the ZPF. This suggests that ZPF-induced forces may be affected through electromagnetic means. As pointed out herein, the ZPF contributes to the passive gravitational energy content of stationary observers simply because the ZPF is anisotropic in the non-inertial frames of those observers. It is not difficult to imagine that some sort of electromagnetic process might be performed that counteracts the gravitationally induced anisotropy of the ZPF. More specifically, were the anisotropy of the ZPF locally annulled or at least substantially reduced within a region in which a stationary observer resides, near a gravitational source, the energy contributed to the observer due to zero-point radiation would be substantially reduced, as well. Assuming that such an alteration can indeed be carried out, it is straightforward to see that in the limit as the ZPF-induced force tends toward zero, the force on the observer, given by Eq. (10), should reduce to a purely gravitational force.

Continuing along these lines, it can be further imagined that through the above-mentioned electromagnetic process the ZPF might be affected to such an extent that the anisotropy of zero-point radiation becomes inverted, thus occurring in a direction opposite to that normally caused by gravitation. The force imparted to the observer by zero-point radiation would then act in opposition to the purely gravitational force. This implies that when the ZPF is altered such that the ZPF-induced force acts upward, the weight of the stationary observer should appear somewhat smaller than when the ZPF is unaltered. Clearly then, as the alteration in the anisotropy of the ZPF is further increased, the observable weight of the stationary observer should be correspondingly decreased. Based on this line of reasoning, when the ZPF is manipulated to such an extent that the resulting ZPF-induced force is equal, but oppositely directed, to the downward gravitation force, the observable weight of the observer should drop to zero.

The next natural question to ask is *how* can such an alteration of the ZPF be carried out? One possible answer may be found in [6] wherein it was shown that when the ZPF performs positive electromagnetic work within a localized region, the energy density of that region appears negative relative to the rest of the universe. More simply stated, when the ZPF performs positive work, the energy density of the ZPF decreases. Based on this, it is not difficult to speculate that any process through which the ZPF performs positive work alters the local field geometry of the ZPF. And with certain field configurations [6], it just may be possible to manipulate the weights of material objects.

As a final thought, it is interesting to ponder the work of Podkletnov and Nieminen [7] wherein a levitating superconducting ring was alleged to have reduced the weights of objects suspended above the ring. It may be that the superconducting ring somehow came into interaction with a portion of the ZPF, causing the ZPF to perform work, and thereby reducing the anisotropy of the ZPF above the ring. If the superconducting ring did indeed bring about a small reduction of the anisotropy of the ZPF, then the weights of objects placed above the ring should appear slightly smaller than when weighed elsewhere. According to Podkletnov and Nieminen [7], such a minute weight-shift was indeed observed; a 5.47834-g sample was observed to loose between 0.05% and 2.0% of its weight when placed above the levitating superconducting ring. It would be interesting to see if this small weight-shift can be derived solely on the basis of the approach presented herein.

## Notes and References